\documentclass[preprintnumbers,amsmath,11pt,amssymb,floatfix,superscriptaddress,showpacs,nofootinbib]{revtex4}
\usepackage{graphicx}
\usepackage{epsfig}
\usepackage{bm}
\usepackage{amsfonts}

\input epsf

\begin{document}

\title{\Large Reconstruction of $f(T)$ gravity from the Holographic dark energy}

\author{Surajit Chattopadhyay}
\email{surajit_2008@yahoo.co.in, surajcha@iucaa.ernet.in}
\affiliation{ Pailan College of Management and Technology, Bengal
Pailan Park, Kolkata, India.}

\author{Antonio Pasqua}
\email{toto.pasqua@gmail.com} \affiliation{Department of Physics,
University of Trieste, Trieste, Italy.}

\date{\today}

\begin{abstract}
Among different candidates to play the role of Dark Energy (DE),
modified gravity has emerged as offering a possible unification of
Dark Matter (DM) and DE. The purpose of this work is to
develop a reconstruction scheme for the modified gravity with
$f(T)$ action using holographic energy density. In the framework
of the said modified gravity we have considered the equation of
state of the Holographic DE (HDE) density. Subsequently we have
developed a reconstruction scheme for modified gravity with $f(T)$
action. Finally we have obtained a modified gravity action
consistent with the HDE scenario.
\end{abstract}


\maketitle

\section{Introduction}
Cosmological observations obtained with Supernovae Ia (SNeIa), the
Cosmic Microwave Background (CMB) radiation anisotropies, the
Large Scale Structure (LSS) and X-ray experiments have well
established the accelerated expansion of our universe
\cite{11,13,14,15,16,17,18}. A missing energy component also known
as Dark Energy (DE) with negative pressure is widely considered by
scientists as responsible of this accelerated expansion. Recent
analysis of cosmological observations
indicates that the two-thirds of the total energy of the universe is been occupied by the DE whereas DM occupies almost the remaining part (the baryonic matter represents only a few percent of the total energy density of the universe) \cite{1-1}. The contribution of the radiation is practically negligible.\\
The nature of DE is still unknown and many candidates have been
proposed in order to describe it  \cite{21,22,23,24,25,26,27,28}.
The simplest one is represented by a tiny positive cosmological
constant, with a negative constant equation of state (EoS)
parameter $\omega$, i.e. $\omega=-1$. However, cosmologists know
that the cosmological constant suffers from two well-known
difficulties, the fine-tuning and the cosmic coincidence problems:
the former asks why the vacuum energy density is so small (of the
order of $10^{-123}$ smaller than what we observe)
and the latter says why vacuum energy and DM are nearly equal today \cite{221,222}.\\
As possible alternative to cosmological constant, dynamical scalar field models have been
proposed some of which are quintessence \cite{31,32,33,34}, phantom \cite{41,42}, f-essence \cite{fff1,fff2} and K-essence \cite{51,52,53}. \\
An important advance in the studies of black hole theory and
string theory is the suggestion of the so called holographic
principle which was proposed by Fischler and Susskind in 1998
\cite{8}. According to the holographic principle, the number of
degrees of freedom of a physical system should be finite and
should scale with its bounding area rather than with its volume
\cite{thooft} and it should be constrained by an infrared cut-off
\cite{12}. The Holographic DE (HDE), based on the holographic
principle, is one of the most studied models of DE
\cite{111,112,113,114,115,116,117,118,119,120,121,122,123}. The HDE
models have also been constrained and tested by various astronomical observations \cite{const1,const2}.\\
Applying the holographic principle to cosmology, the upper bound of the entropy contained
in the universe can be obtained. Following this line, Li \cite{10} suggested as constraint on the energy density of the universe $\rho_\Lambda\leq 3\gamma M^2_pL^{-2}$, where $\gamma$ is a numerical constant,
$L$ is the IR cut-off radius and $M_p$ is the reduced Planck mass. The equality sign holds when the holographic bound is saturated.\\
Importance of modified gravity for late acceleration of the
universe has been reviewed \cite{otto1,otto2}. Various modified
gravity theories have been proposed so far: some of the most
studied include $f\left( R \right)$ \cite{dieci},  $f\left( G
\right)$ \cite{sedici1,sedici2}, Horava-Lifshitz \cite{diciotto}
and Gauss-Bonnet \cite{venti} theories.  Recently, a new theory of
gravity known as
$f\left( T \right)$ gravity, which is formulated in a space-time possessing absolute parallelism \cite{dodici1,dodici2}, has been proposed.\\
Fundamental aspects of $f\left( T \right)$ gravity have been
recently studied \cite{ventidue1,ventidue2}. In the  $f\left( T
\right)$ theory of gravity, the teleparallel Lagrangian density
described by the torsion scalar $T$ has been promoted to be a
function of $T$, i.e. $f\left( T \right)$,  in order to account
for the late time cosmic acceleration \cite{ventiquattro1}. In a
recent work, Jamil et al \cite{jam} examined the interacting DE
model in $f\left( T \right)$ cosmology assuming DE as a perfect
fluid and choosing a specific cosmologically viable form $f\left(
T \right) = \beta \sqrt{T}$. Statefinder diagnostic of $f(T)$
gravity has been studied in \cite{prabir}. Purpose of the present
work is to develop a reconstruction scheme for the modified
gravity with $f(T)$ action
using holographic energy density.\\

\section{Reconstruction of $f(T)$ gravity}
In the framework of $f(T)$ theory, the action of modified
teleparallel action is given by \cite{myrza3}:
\begin{equation}
I=\frac{1}{16\pi G}\int d^{4}x \sqrt{-g}\left[f(T)+L_{m}\right],
\label{1}
\end{equation}
where $L_{m}$ is the Lagrangian density of the matter inside the
universe, $G$ is the gravitational constant and $g$ is the determinant of the metric tensor $g^{\mu \nu}$. We consider a flat Friedmann-Robertson-Walker (FRW)
universe filled with the pressureless matter. Choosing $(8\pi
G=1)$, the modified Friedman equations in the framework of $f(T)$
gravity are given by \cite{Ferraro, myrza3}:
\begin{eqnarray}
H^{2}&=&\frac{1}{3}\left(\rho+\rho_{T}\right), \label{2}\\
2\dot{H}+3H^{2}&=&-\left(p+p_{T}\right), \label{3}
\end{eqnarray}
where
\begin{eqnarray}
\rho_{T}&=&\frac{1}{2}(2T f_{T}-f-T), \label{4}\\
p_{T}&=&-\frac{1}{2}\left[-8\dot{H}T
f_{TT}+(2T-4\dot{H})f_{T}-f+4\dot{H}-T\right],\label{5}
\end{eqnarray}
and \cite{Ferraro}:
\begin{equation}
T=-6\left(H^{2}\right). \label{6}
\end{equation}
Earlier, Setare \cite{Setare1} reconstructed $f(R)$ gravity from HDE. In this Section, we shall discuss a
reconstruction of $f(T)$ gravity in HDE scenario. We must emphasize that reconstruction of $f\left(T\right)$ gravity has been already studied in \cite{odi1} and a general DE review which discusses the reconstruction in $f\left(T\right)$ gravity can be found in \cite{odi2}.\\
Following \cite{Setare1}, the HDE density is chosen as:
\begin{equation}
\rho_{\Lambda}=\frac{3 c^{2}}{R_{h}^{2}}, \label{7}
\end{equation}
where $R_{h}$ represents the future event horizon and $c$ is a constant.\\
The expression of $R_{h}$ is given by:
\begin{equation}
R_{h}=a\int_{t}^{\infty}\frac{dt}{a}=a\int_{a}^{\infty}\frac{da}{Ha^{2}}. \label{8}
\end{equation}
The dimensionless DE is defined by using the critical energy density $\rho_{cr}=3H^{2}$ as follow:
\begin{equation}
\Omega_{\Lambda}=\frac{\rho_{\Lambda}}{\rho_{cr}}=\frac{c^{2}}{R_{h}^{2}H^{2}}. \label{9}
\end{equation}
The time derivative of the future horizon is given by:
\begin{equation}
\dot{R}_{h}=R_{h}H-1=\frac{c}{\sqrt{\Omega_{\Lambda}}}-1. \label{10}
\end{equation}
Using the conservation equation, the EoS parameter for HDE has been obtained by \cite{Setare1} as follow:
\begin{equation}
\omega_{\Lambda}=-\left(\frac{1}{3}+\frac{2\sqrt{\Omega_{\Lambda}}}{3c}\right). \label{11}
\end{equation}
In order to reconstruct $f(T)$ gravity in HDE scenario, we replace the energy density of Eq. (\ref{2}) with $\rho_{\Lambda}$ and hence we get:
\begin{equation}
\rho_{\Lambda}=6H^{2}f_{T}+\frac{1}{2}f(T). \label{12}
\end{equation}
Using Eq. (\ref{9}) in Eq. (\ref{12}), we can write:
\begin{equation}
6H^{2}\Omega_{\Lambda}=12H^{2}f_{T}+f(T). \label{13}
\end{equation}
Hence, we get
\begin{equation}
f(T)=6H^{2}\Omega_{\Lambda}-12H^{2}f_{T}=-T(\Omega_{\Lambda}-2f_{T}). \label{14}
\end{equation}
We now consider Eq. (\ref{3}), where $p$ would be replaced by $p_{\Lambda}$. As $\omega_{\Lambda}=p_{\Lambda}/\rho_{\Lambda}$, we can write
\begin{equation}
p_{\Lambda}=3H^{2}w_{\Lambda}\Omega_{\Lambda}, \label{15}
\end{equation}
where $\omega_{\Lambda}$ is given in Eq. (\ref{11}). Using Eqs. (\ref{3}), (\ref{5}) and (\ref{15}), we get:
\begin{equation}
3\omega_{\Lambda}H^{2}\Omega_{\Lambda}=-4\dot{H}Tf_{TT}+(T-2\dot{H})f_{T}-\frac{1}{2}f(T). \label{16}
\end{equation}
Subsequently, using $T=-6H^{2}$, we can write:
\begin{equation}
3\omega_{\Lambda}\Omega_{\Lambda}=24\dot{H}f_{TT}-\left(6+\frac{2\dot{H}}{H^{2}}\right)f_{T}-\frac{1}{2H^{2}}f(T). \label{17}
\end{equation}
For scale factor $a\left(t\right)$ we shall consider the solution \cite{Setare1}:
\begin{equation}
a(t)=a_{0}(t_{s}-t)^{n}, \label{18}
\end{equation}
where, $a_{0}$, $t_{s}$ and $n$ are constants. Hence:
\begin{eqnarray}
H &=& -\frac{n}{t_{s}-t}, \\
T &=& -\frac{6n^{2}}{(t_{s}-t)^{2}}, \\
\Omega_{\Lambda} &=& c^{2}\left(1-\frac{1}{n}\right)^{2}. \label{19}
\end{eqnarray}
The solution of Eq. (\ref{14}) is given by:
\begin{equation}
f(T)=\left[\frac{c(n-1)}{n}\right]^{2}(T-2)+C_{1}e^{-\frac{T}{2}}, \label{20}
\end{equation}
where $C_1$ is a constant.\\
Instead, the solution of Eq. (\ref{17}) is given by:
\begin{equation}
f(T)=\frac{\alpha}{\beta-3}T+T^{\frac{\beta_{2}-\beta_{1}-\sqrt{(\beta_{1}-\beta_{2})^{2}+12\beta_{2}}}{2\beta_{2}}}\left(T^{\frac{\sqrt{(\beta_{1}-\beta_{2})^{2}+12\beta_{2}}}{\beta_{2}}}C_{2}+C_{3}\right), \label{21}
\end{equation}
where
\begin{eqnarray}
\alpha &=& \frac{c^{2}(n-1)^{2}(2-3n)}{n^{3}}, \\
\beta_{1} &=& -6+\frac{2}{n}, \\
\beta_{2} &=& \frac{4}{n}, \label{22}
\end{eqnarray}
and $C_2$ and $C_3$ are two constants.

\begin{figure}[h]
\begin{minipage}{16pc}
\includegraphics[width=16pc]{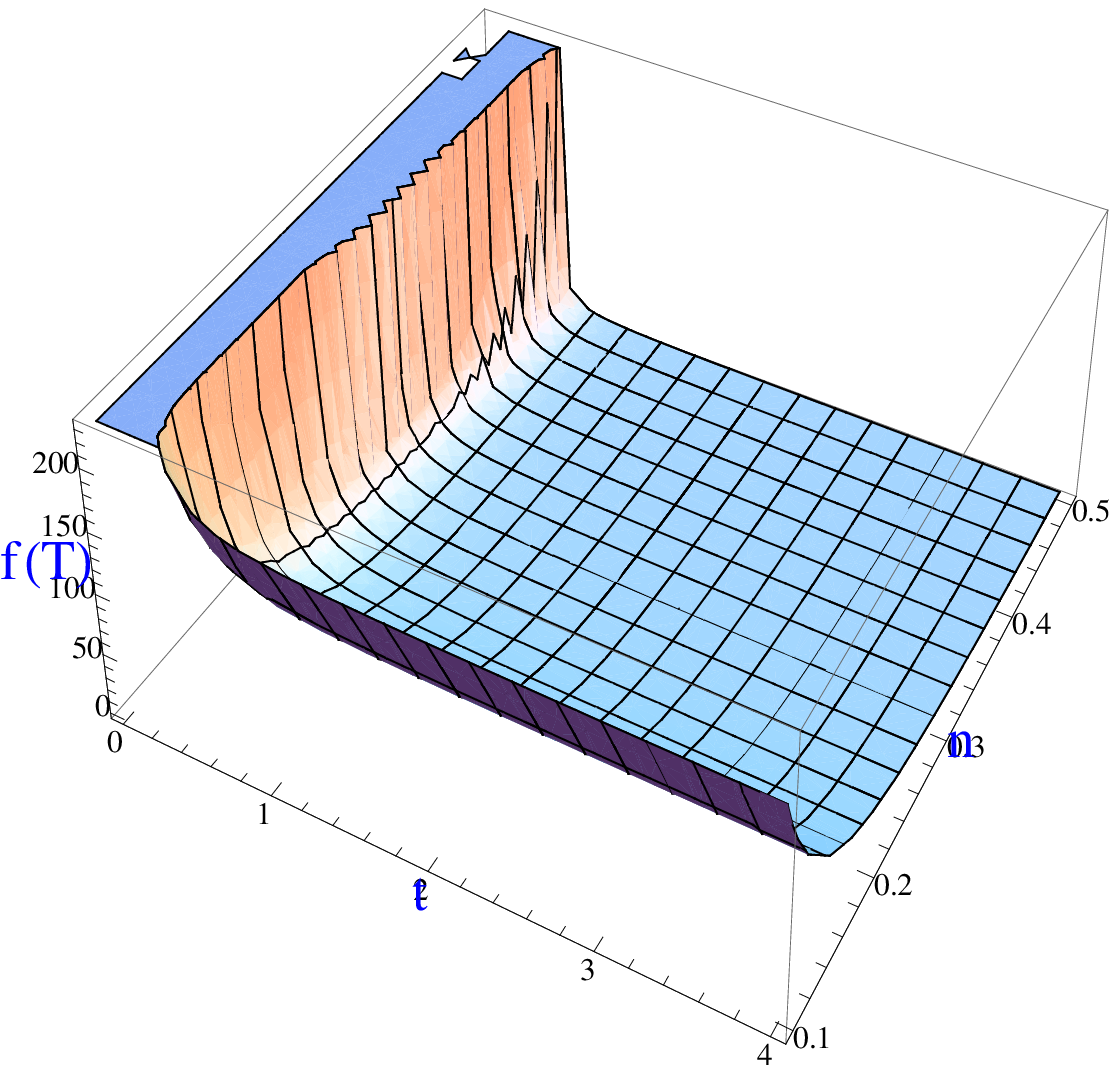}
\caption{\label{label}This Figure plots evolution of reconstructed
$f(T)$  for solution given in Eq. (\ref{20}) and we find $f(T)\geq 0$ with the
evolution of the universe for a range of $n$. The $x$-axis plots
$t$ and $y$-axis plots $n$.}
\end{minipage}\hspace{3pc}%
\begin{minipage}{16pc}
\includegraphics[width=16pc]{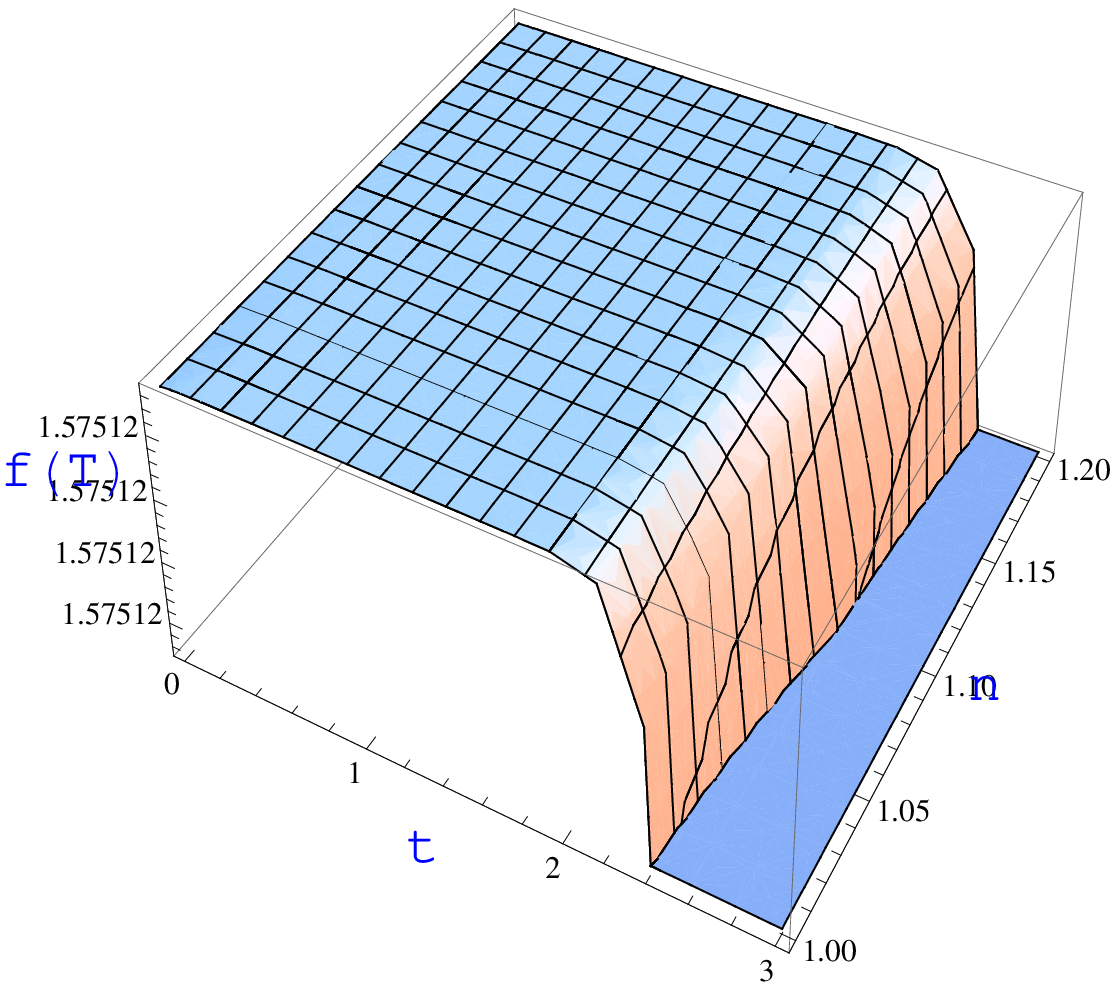}
\caption{\label{label}This Figure plots evolution of reconstructed
$f(T)$ for solution given in Eq. (\ref{21}) and we find $f(T)\geq 0$ with the
evolution of the universe for a range of $n$. The $x$-axis plots
$t$ and $y$-axis plots $n$.}
\end{minipage}\hspace{3pc}%
\end{figure}

\begin{figure}[h]
\begin{minipage}{16pc}
\includegraphics[width=16pc]{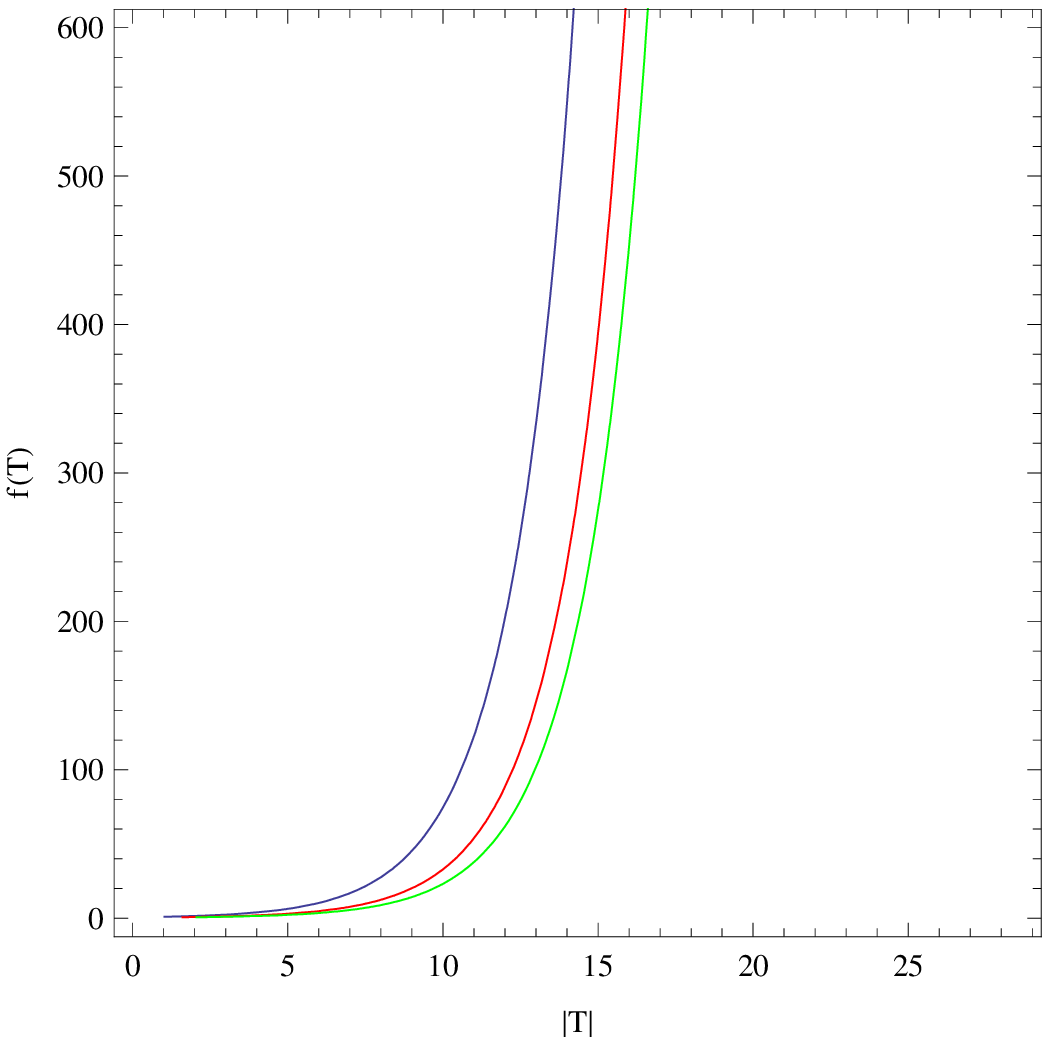}
\caption{\label{label}This Figure plots $f(T)$  for solution given in Eq. (\ref{20})
against $|T|$ for different values of $n$.}
\end{minipage}\hspace{3pc}%
\begin{minipage}{16pc}
\includegraphics[width=16pc]{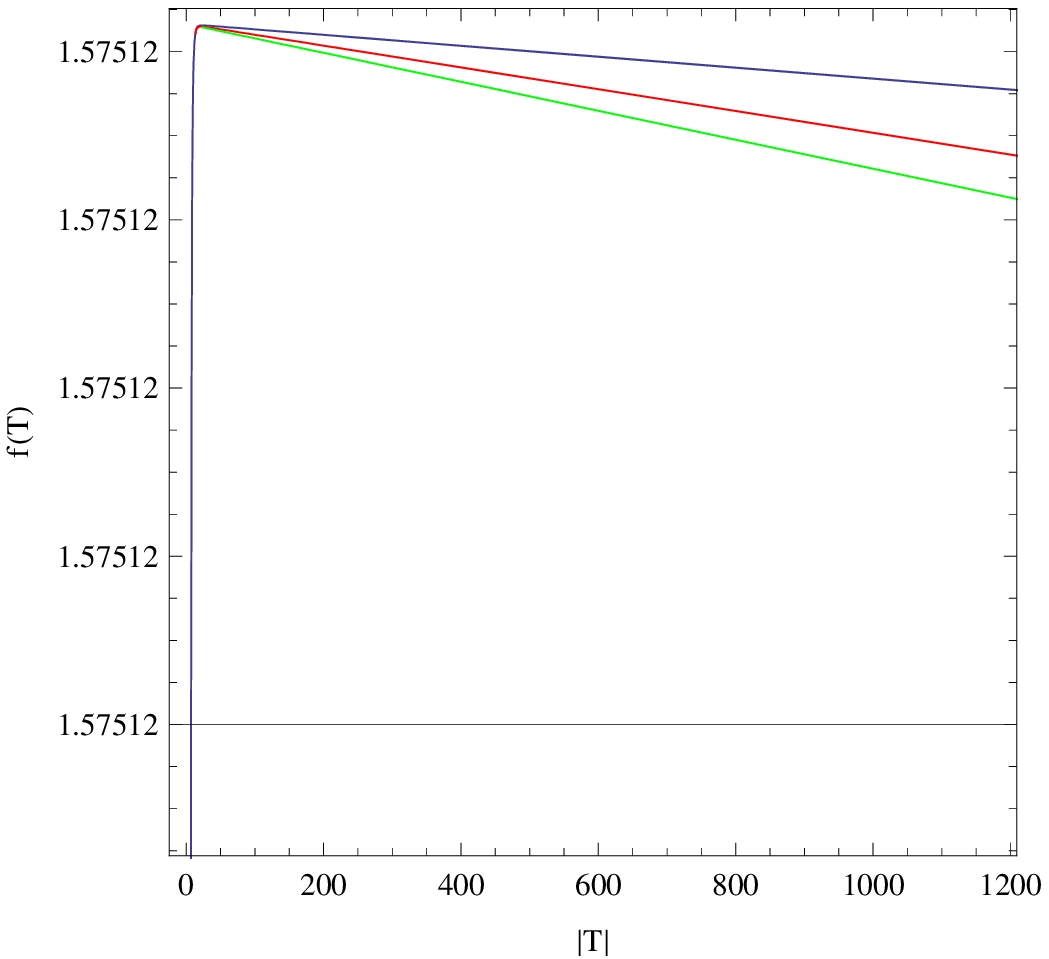}
\caption{\label{label}This Figure plots $f(T)$  for solution given in Eq. (\ref{21})
against $|T|$ for different values of $n$.}
\end{minipage}\hspace{3pc}%
\end{figure}
The solutions for $f(T)$ are now studied graphically. In Figure 1,
we plot the solution $f(T)$ given in Eq. (\ref{20}) against $t$ and $n$. The three
dimensional plot shows a decaying behavior of $f(T)$ with
evolution of the universe. Furthermore, the rate of decay is
increasing with increase in the value of $n$. Similarly, when we
plot the expression of $f(T)$ given in (\ref{21}) in Figure 2, we find a decaying pattern of $f(T)$ with
passage of cosmic time. However, contrary to what happened in the
earlier case, the rate of decay is less in the case of higher
values of $n$. In Figures 3 and 4 we have viewed the behavior of
$f(T)$ with variation in $T$. We observe that for the solution
corresponding to Eq. (\ref{20}), the $f(T)$ increases with increase
in absolute value of $T$. On the other hand, the $f(T)$ obtained
from Eq. (\ref{21}) decays with $|T|$. It has been discussed in reference
\cite{rastkar} that satisfaction of the above condition is a
sufficient condition for a realistic model. From Figure 3, we
understand that $f(T)\rightarrow 0$ as $T\rightarrow 0$ for the
solution obtained from Eq. (\ref{20}). However, this does not occur for the
solution obtained from Eq. (\ref{21}). Thus, it may be stated that the
solution obtained in Eq. (\ref{20}) is a more realistic model than that
obtained in Eq. (\ref{21}).\\

\section{Conclusion}
Among different candidates to play the role of the DE,
modified gravity has emerged as a possible unification of
DM and DE. The present work aims at a
cosmological application of the HDE density in the
framework of a modified gravity, named as $f(T)$ gravity. In the
framework of the said modified gravity, we have considered the
equation of state of the HDE density.
Subsequently, we have developed a reconstruction scheme for
modified gravity with $f(T)$ action. Considering De
density given in Eq. (4) in holographic form and then by assuming a
simple solution for the scale factor $a\left(t \right)$ as given in Eq. (\ref{18}), we have obtained a solution for
differential equation for $f(T)$ in Eq. (\ref{20}). Again, by
considering the DE pressure in holographic form, we have
created a differential equation for $f(T)$ in Eq. (\ref{17}) and using the
same simple solution for the scale factor, we derived $f(T)$. It has been
revealed that the solution of Eq. (\ref{20}) seems to be more realistic than
that of Eq. (\ref{21}). In this, way we get a modified gravity action
consistent with the HDE scenario.
\\

\subsection{Acknowledgements}
The first author sincerely acknowledges the facilities provided to
him by the Inter-University Centre for Astronomy and Astrophysics
(IUCAA), Pune, India, during his visit in November, 2012 under the
Visitor Associateship Programme.
\\

\end{document}